\documentstyle[11pt,newpasp,twoside,epsf]{article}
\def\plotone#1{\centering \leavevmode\epsfxsize=0.55\columnwidth \epsfbox{#1}}

\newcommand{\ion}[2]{{\rm #1}\,{\small\rm #2}}
\newcommand{\HI}{\ion{H}{I}}
\newcommand{\HeII}{\ion{He}{II}}
\newcommand{\HeIII}{\ion{He}{III}}
\newcommand{\SiIV}{\ion{Si}{IV}}
\newcommand{\CIV}{\ion{C}{IV}}
\newcommand{\lya}{Ly$\alpha$}

\begin{document}

\title{Constraining reionization using the thermal history of the baryons}
\author{Joop Schaye}
\affil{Institute of Astronomy, Madingley Road, Cambridge CB3~0HA, UK}
\author{Tom Theuns}
\affil{Max-Planck-Institut f\"ur Astrophysik, Postfach 1523, 85740
 Garching, Germany}
\author{Michael Rauch}
\affil{European Southern Observatory, Karl-Schwarzschild-Str.\ 2,
85748 Garching, Germany}
\author{George Efstathiou}
\affil{Institute of Astronomy, Madingley Road, Cambridge CB3~0HA, UK}
\author{Wallace L. W. Sargent}
\affil{Astronomy Department, California Institute of Technology,
Pasadena, CA 91125, USA}

\begin{abstract}
The thermal evolution of the intergalactic medium (IGM) depends on the
reionization history of the universe. Numerical simulations indicate
that the low density IGM, which is responsible for the low column
density \lya\ forest, follows a well defined temperature-density
relation. This results in a cut-off in the distribution of line widths
as a function of column density. We use hydrodynamic simulations to
calibrate the relation between the cut-off and the temperature-density
relation and apply this relation to Keck spectra spanning a
redshift range $z=2$--4.5. We find that the temperature peaks at
$z\sim 3$ and interpret this as evidence for reheating due to the
reionization of helium.
\end{abstract}


\section{Introduction}
Quasars have provided us with a unique probe of the high redshift
universe. These bright point sources shine like a flashlight through
space, revealing the presence of baryonic matter through the light it
absorbs. Thus every quasar spectrum contains a one-dimensional map of
the distribution of matter along the line of sight. The extraordinary
quality of the spectra obtainable with the HIRES spectrograph on the
Keck telescope, enables us to extract the wealth of information that
has been collected by the quasar's light along its journey through
space and time. Computer simulations of structure formation have been
remarkably successful in reproducing these observations. They show
that the physics governing the high redshift intergalactic medium
(IGM), which is responsible for the low column density absorption
lines (the so-called \lya\ forest) is relatively simple. The IGM,
which contains most of the baryons in the universe, is photoionized
and photoheated by the collective UV radiation from young stars and
quasars. On large scales its dynamics are determined by the
gravitational field of the dark matter, while on small scales gas
pressure is important. The availability of superb data and a detailed
physical model, have made the \lya\ forest into a powerful probe of
the high-redshift universe.

Since shock heating is unimportant in the low-density IGM, most of the
gas follows a simple temperature-density relation which is the result
of the interplay of photoionization heating and adiabatic cooling due
to the expansion of the universe. For densities around the cosmic
mean, this relation is well-described by a power-law,
$T=T_0(\rho/\bar{\rho})^{\gamma-1}$ (Hui \& Gnedin 1997). At
reionization the gas is reheated, resulting in an increase in $T_0$
and a decrease in $\gamma$ (provided that the gas is reionized on a
timescale short compared to the Hubble time). In ionization
equilibrium, $T_0$ decreases and the slope of the effective equation
of state steepens (i.e.\ $\gamma$ increases). However, because the
timescale for recombination is long, the gas retains some memory of
how and when it was reionized (Miralda-Escud\'e \& Rees 1994).

The distribution of line widths ($b$-parameters) depends on various
mechanisms. Thermal motions of the hydrogen atoms broaden
the \HI\ absorption lines and other processes, such as the differential
Hubble flow across the absorbing structure and bulk flows, also
contribute to the line widths. However, the minimum line width is set
by the temperature of the gas, which in turn depends on the density.
A standard way of analyzing \lya\ forest spectra is by decomposing
them into a set of Voigt profiles. Since the minimum line width
($b$-parameter) depends on the temperature, and since column density
($N$) correlates strongly with physical density, there is a cut-off in
the $b(N)$ distribution which traces the effective equation of state
of the IGM (Schaye et al.~1999; Ricotti, Gnedin \& Shull 2000; Bryan
\& Machacek 2000). We have used this relation to measure
the thermal evolution of the IGM from a set of nine \lya\ quasar
absorption line spectra.

This work is more fully described and discussed in a forthcoming
publication (Schaye et al.\ 2000).

\section{Method}
We have measured the $b(N)$ cut-off for a set of nine high-quality
\lya\ forest spectra, spanning the redshift range 2.0--4.5, eight of
which were taken with the HIRES spectrograph of the Keck telescope.
We used hydrodynamic simulations to calibrate the relations between
the $b(N)$ cut-off and the temperature-density relation. Except for
the two lowest redshift quasars, the \lya\ forest spectra were split
in two in order to take into account the significant redshift
evolution ($\Delta z \sim 0.5$) and signal-to-noise variation across
a single spectrum. The calibration was done separately for each half
of each observed spectrum. The synthetic spectra were processed to
give them identical characteristics (resolution, pixel size, noise
properties, mean absorption) as the real data. The same Voigt profile
fitting package (an automated version of VPFIT (Webb 1987)) was used
for both the simulated and the observed spectra.

\section{Results and discussion}
The measured evolution of the temperature at the mean density and the
slope of the effective equation of state are plotted in Figure~1. From
$z\sim 3.5$ to $z\sim 3.0$, $T_0$ increases and the gas 
becomes close to isothermal ($\gamma \sim 1.0$). This behavior differs
drastically from that predicted by models in which helium is fully
reionized at higher redshift. For example, the solid lines correspond
to a simulation that uses a uniform metagalactic UV-background from
quasars as computed by Haardt \& Madau (1996) and which assumes the gas to
be optically thin. In this simulation, both hydrogen and helium are
fully reionized by $z\sim 4.5$ and the temperature of the IGM declines
slowly as the universe expands. Such a model can clearly not account
for the peak in the temperature at $z\sim 3$ (reduced $\chi^2$ for the
solid curves are 6.9 for $T_0$ and 3.6 for $\gamma$). Instead, we
associate the peak in $T_0$ and the low value of $\gamma$ with reheating
due to the second reionization of helium (\HeII\ $\rightarrow$
\HeIII). This interpretation is supported by measurements of the
\SiIV/\CIV\ ratio (Songaila 1998, but see also Boksenberg, Rauch, \&
Sargent 1998 and Giroux \& Shull 1997) and direct measurements of the
\HeII\ opacity (Heap et al.\ 2000 and references therein).

\begin{figure}
\mbox{
\hspace{-0.8cm}
\plotone{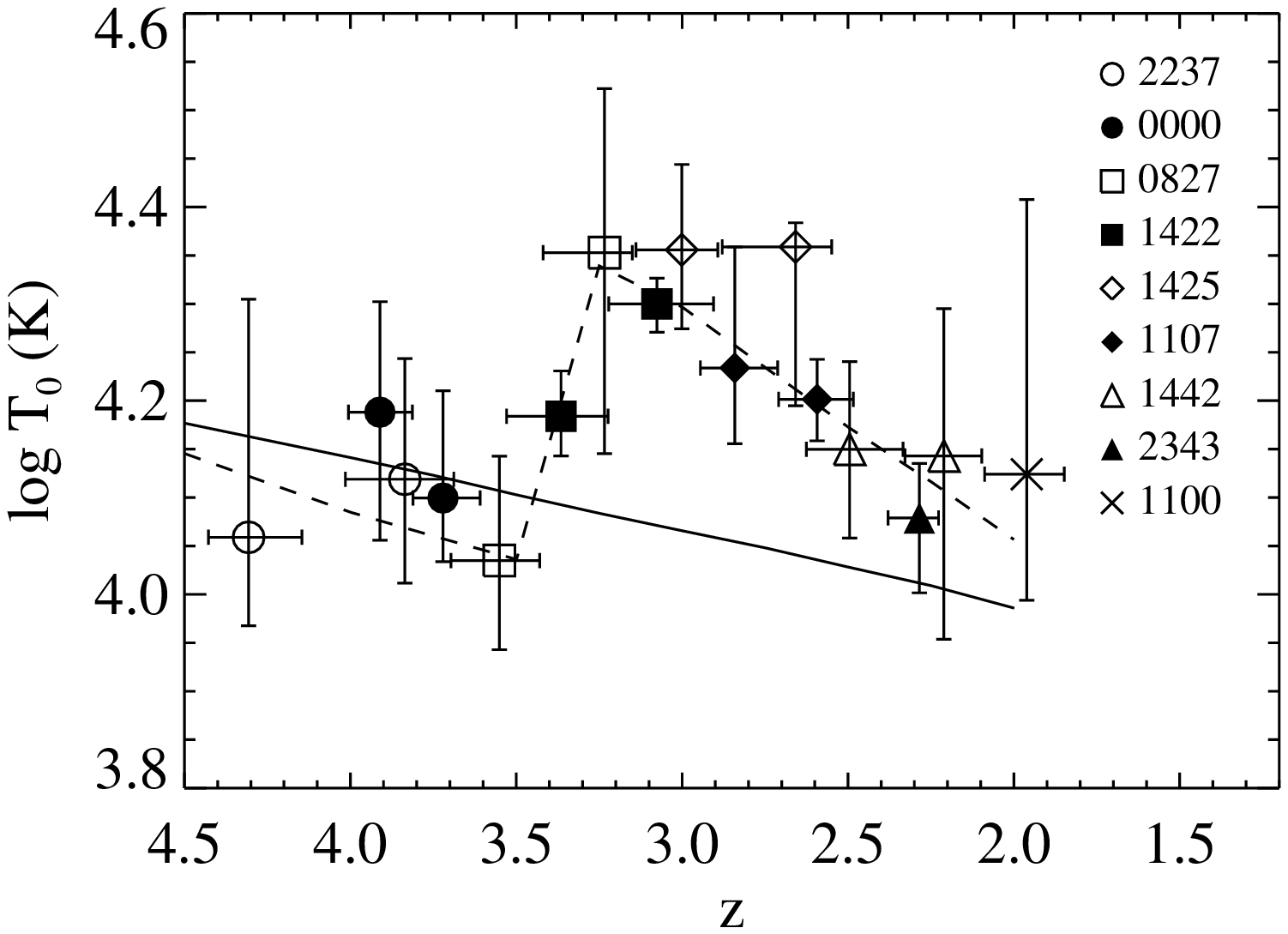}
\hspace{-0.8cm}
\plotone{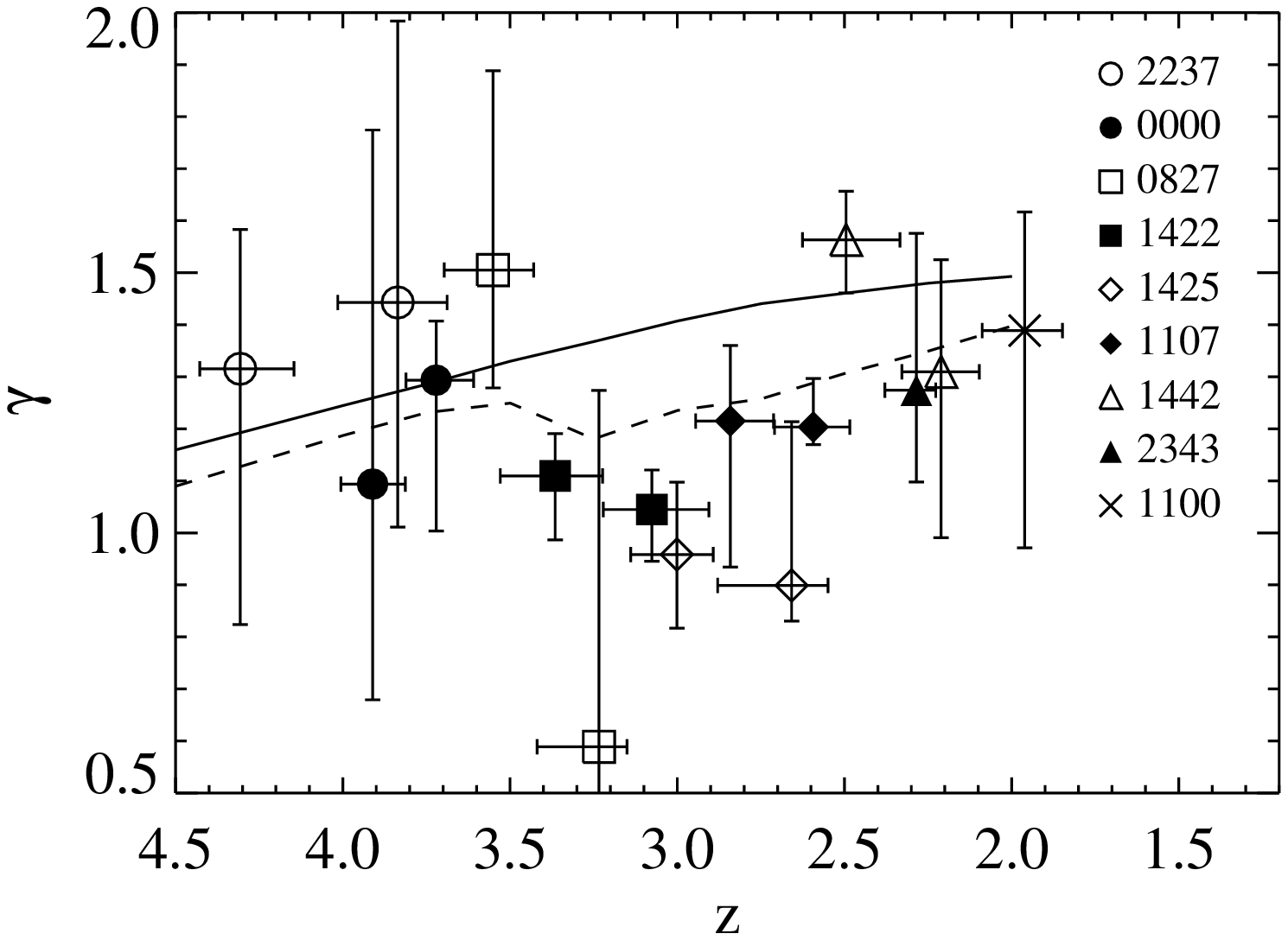} 
}
\caption{The evolution of the temperature at the mean density (left
panel) and the slope of the effective equation of state (right
panel). Horizontal error bars indicate redshift interval spanned by
the absorption lines, vertical error bars are $1\,\sigma$
errors. Different symbols correspond to different quasars. See text
for a description of the models.}
\end{figure}

The dashed lines in Figure~1 are for a model that was designed to fit
the data (reduced $\chi^2$ is 0.24 for $T_0$ and 1.38 for
$\gamma$). In this simulation, which has a much softer UV-background
at high redshift, \HeII\ reionizes at $z\sim 3.2$. Before
reionization, when the gas is optically thick to ionizing photons, the
mean energy per photoionization is much higher than in the optically
thin limit (Abel \& Haehnelt 1999). We have approximated this effect
in this simulation by enhancing the photoheating rates during
reionization, so raising the temperature of the IGM.

Since the simulation assumes a uniform ionizing background, the
temperature has to increase abruptly (i.e.\ much faster than the gas
can recombine) in order to make $\gamma$ as small as observed. In
reality, the low-density gas may be reionized by harder photons, which
will be the first ionizing photons to escape from the dense regions
surrounding the sources. This would lead to a larger temperature
increase in the more dilute, cooler regions, resulting in a decrease
of $\gamma$ even for a more gradual reionization. Furthermore,
although reionization may proceed fast locally (as in our small
simulation box), it may be patchy and take some time to
complete. Hence the steep temperature jump indicated by the dashed
line, although compatible with the data, should be regarded as
illustrative only. The globally averaged $T_0$ could well increase
more gradually which would also be consistent with the data. More data
at $z \ga 3$ is needed to determine whether the temperature rise is
sharp or gradual. On the theoretical side, more realistic models
should include radiative transfer effects, which are important during
reionization.

Together with measurements of the \HeII\ opacity, which probe the
ionization state in the voids, the thermal history of the IGM provides
important constraints on models of helium reionization. Furthermore,
the temperature of the IGM before the onset of helium reionization can
be used to constrain the redshift of hydrogen reionization, which
marks the end of the dark ages of cosmic history.

\acknowledgments
We are grateful to Bob Carswell and Sara Ellison for giving us
permission to use their spectra of the quasars Q1100$-$264 and
APM\,08279+5255 respectively.

\end{document}